\documentclass[aps.prb,amsmath,amssymb,twocolumn,floatfix]{revtex4-2}
\usepackage{graphicx}
\usepackage{dcolumn}
\usepackage{bm}

\begin{document}

\title{Absence of Spontaneous Magnetic Fields Due to Time-Reversal Symmetry Breaking in Bulk Superconducting UTe$_2$}
\author{N.~Azari,$^1$ M.~Yakovlev,$^1$ N.~Rye,$^1$ S. R.~Dunsiger,$^{1,2}$ S.~Sundar,$^3$ M. M.~Bordelon,$^4$ S. M.~Thomas,$^4$ J. D. Thompson,$^4$ P. F. S.~Rosa,$^4$ and J. E.~Sonier$^1$}

\affiliation{$^1$Department of Physics, Simon Fraser University, Burnaby, British Columbia V5A 1S6, Canada \\
$^2$Centre for Molecular and Materials Science, TRIUMF, Vancouver, British Columbia V6T 2A3, Canada \\
$^3$Scottish Universities Physics Alliance, School of Physics and Astronomy, University of St. Andrews, St. Andrews KY16 9SS, United Kingdom \\
$^4$Los Alamos National Laboratory, Los Alamos, New Mexico 87545, USA}  

\date{\today}
\begin{abstract}
We have investigated the low-temperature local magnetic properties in the bulk of molten salt-flux (MSF) grown single crystals of the candidate odd-parity 
superconductor UTe$_2$ by zero-field muon spin relaxation ($\mu$SR). In contrast to previous $\mu$SR studies of UTe$_2$ single crystals grown by a chemical
vapour transport (CVT) method, we find no evidence of magnetic clusters or electronic moments fluctuating slow enough to cause a discernible relaxation of the
zero-field $\mu$SR asymmetry spectrum. Consequently, our measurements on MSF-grown single crystals rule out the generation of
spontaneous magnetic fields in the bulk that would occur near impurities or lattice defects if the superconducting state of UTe$_2$ breaks time-reversal symmetry.
This result suggests UTe$_2$ is characterized by a single-component superconducting order parameter.
\end{abstract}
\maketitle
A superconducting phase results from the condensation of electron (Cooper) pairs into a coherent
quantum state characterized by a pair wavefunction or complex order parameter consisting of an amplitude and a phase.
All superconductors break gauge symmetry, which 
means their order parameter adopts a well defined phase below the superconducting transition temperature ($T_c$).
Conventional superconductors have a spin-singlet pairing state with an isotropic spatial component ($s$-wave) mediated by electron-phonon coupling.  
By contrast, unconventional superconductors deviate from this pairing state and may be characterized by more elaborate superconducting order
parameters due to the breaking of additional symmetries \cite{Sigrist:91}.

Recently-discovered UTe$_2$ is believed to be a rare example of an unconventional odd-parity superconductor based on an abnormally large upper critical magnetic
field ($H_{c2}$) \cite{Ran:19,Aoki:19}, as well as the small change in the nuclear-magnetic resonance (NMR) Knight shift as a function of temperature 
in the superconducting state \cite{Nakamine:19,Fujibayashi:22,Matsumura:23}.   
The observation of a nonzero polar Kerr effect (PKE) in the superconducting state of CVT-grown UTe$_2$ single crystals that exhibit two phase transitions in the specific heat
has been taken as evidence of a time-reversal symmetry (TRS) breaking order parameter \cite{Hayes:21,Wei:22}.
Broken TRS is a defining property of chiral superconductivity \cite{Kallin:16}, and hence the appearance of the PKE below $T_c$ lends support
to other signatures of chiral superconductivity detected by scanning tunneling microscopy (STM) \cite{Jiao:20} and magnetic penetration depth \cite{Bae:21,Ishihara:23} measurements
of UTe$_2$. Chiral odd-parity superconductivity is of much current interest, because certain topological non-trivial Cooper pairing states can host Majorana zero modes 
with potential applications for topological quantum computing \cite{Kallin:16,Sarma:15}.

The superconducting order parameter at a second order phase transition is restricted to an irreducible representation of the total symmetry group \cite{Sigrist:91}. 
The possible superconducting phases in UTe$_2$ are therefore classified by the crystalline point group symmetry $D_{2h}$. Order parameters that
transform under the one-dimensional irreducible representations (1-D irreps) of this group do not break TRS. Consequently, to explain the signature of spontaneous TRS breaking
in the polar Kerr measurements and the chiral surface states detected by STM, an  
odd-parity superconducting order parameter having two nearly degenerate components with a relative phase belonging to different 1-D irreps of the $D_{2h}$ crystalline 
point group has been proposed \cite{Hayes:21,Jiao:20}. However, this characterization of the superconducting order parameter for UTe$_2$ presents a number of challenges.
The PKE was observed in a sample showing two phase transitions in the specific heat at ambient pressure, as expected if the two components of the order parameter 
are nearly degenerate. However, as the quality of the samples improved, only a single superconducting transition was observed \cite{Rosa:22,Sakai:22}. 
A recent study of the PKE in CVT-grown and MSF-grown UTe$_2$ single crystals that exhibit a single superconducting phase transition in the specific heat
found no evidence for TRS breaking superconductivity \cite{Ajeesh:23}. Furthermore, pulse-echo ultrasound measurments of the changes in elastic moduli across $T_c$ in single and double
phase transition CVT-grown samples \cite{Theuss:23} and recent NMR Knight shift measurements on MSF-grown UTe$_2$ single crystals \cite{Matsumura:23} both
favor a single-component odd parity superconducting order parameter.

Zero-field muon spin relaxation (ZF-$\mu$SR) is an ideal tool for independently determining whether TRS symmetry is spontaneoulsy broken in a superconducting state.
In the bulk of a TRS breaking superconductor, inhomogeneities of the order parameter that occur near impurities, lattice defects, or around domain walls
generate spontaneous currents \cite{Sigrist:98}. The corresponding weak spontaneous local magnetic fields have been detected by ZF-$\mu$SR in numerous
unconventional superconductors --- most notably UPt$_3$ \cite{Luke:93} and Sr$_2$RuO$_4$ \cite{Luke:98}, two compounds in which TRS breaking has
been confirmed by polar Kerr measurements \cite{Xia:06,Schemm:14}. Both superconductors have been considered as likely chiral superconductors, although
Sr$_2$RuO$_4$ is no longer believed to be a candidate for odd-parity superconductivity \cite{Chronister:21}.  

As for UTe$_2$, the relaxation rate of 
the ZF-$\mu$SR signal from CVT-grown single crystals exhibiting single or double transitions in the specific heat was found to be dominated by inhomogeneous
freezing of magnetic clusters \cite{Sundar:19,Sundar:23}, which thwarted sensitivity to the potential onset of weak spontaneous fields at $T_c$.
As discussed in Ref.~\cite{Sundar:23}, the magnetic clusters are likely responsible for the residual linear term in the temperature dependence of the specific heat ($C$) below $T_c$
and the low-temperature upturn in $C/T$ versus $T$ that are ubiquitous in UTe$_2$ samples grown via the CVT method. Moreover, a saturation in the growth of the total volume of the 
magnetic clusters and an abrupt slowing down of their fluctuation rate was observed near $T_c$. 
By contrast, current UTe$_2$ single crystals grown by the MSF method have less disorder,
and correspondingly a much larger residual resistivity ratio (RRR) as well as a substantially smaller residual $T$-linear term in the specific heat compared 
to CVT-grown single crystals \cite{Sakai:22}. A potential origin of the magnetic clusters is discussed later, but they are presumably induced by defects that disrupt long-range electronic 
correlations, as suggested in Ref.~\cite{Tokunaga:22}. Hence while MSF-grown UTe$_2$ single crystals may contain trace 
amounts of the ferromagnetic (FM) impurities U$_7$Te$_{12}$ and U$_3$Te$_5$ \cite{Sakai:22}, magnetic clusters are expected to be sparse.
\begin{figure}
\centering
\includegraphics[width=\columnwidth]{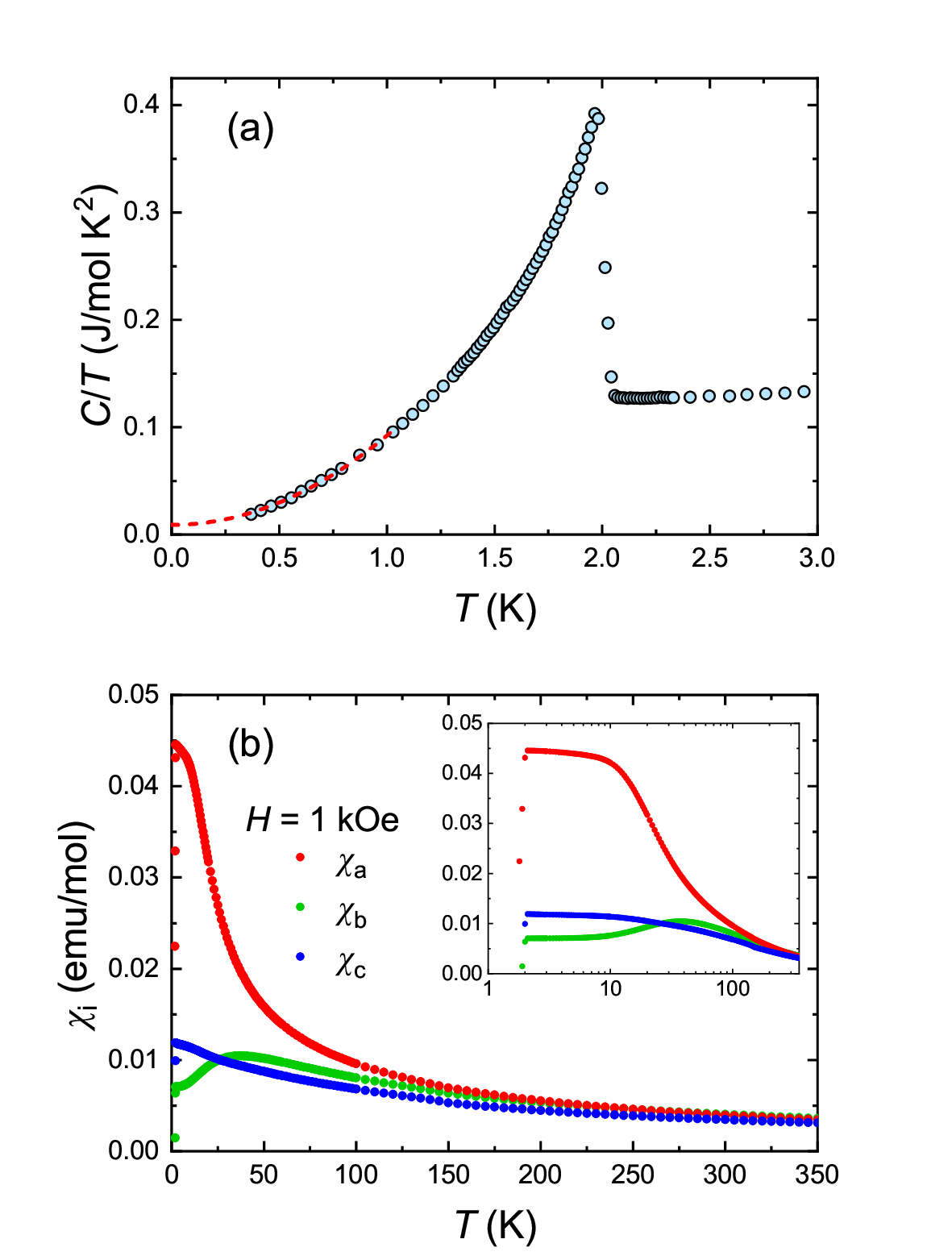}
\caption{(a) Temperature dependence of the specific heat ($C$) for one of the MSF-grown UTe$_2$ single crystals, plotted as $C/T$ vs. $T$.
The dashed curve is a fit of the data below $T \! = \! 1$~K to $C/T \! = \! \gamma^* \! + \! \beta T^2$, which yields $\gamma^* \! = \! 9.1(6)$~mJ/mol$\cdot$K$^2$
and $\beta \! = \! 83(1)$~mJ/mol$\cdot$K$^3$.
(b) Temperature dependence of the bulk magnetic susceptibility of the single crystal for a magnetic field $H \! = \! 1$~kOe applied parallel to the three different
principal crystallographic axes. The inset shows the same data as a semi-logarithmic plot.}
\label{fig1}
\end{figure}  
\begin{figure}
\centering
\includegraphics[width=\columnwidth]{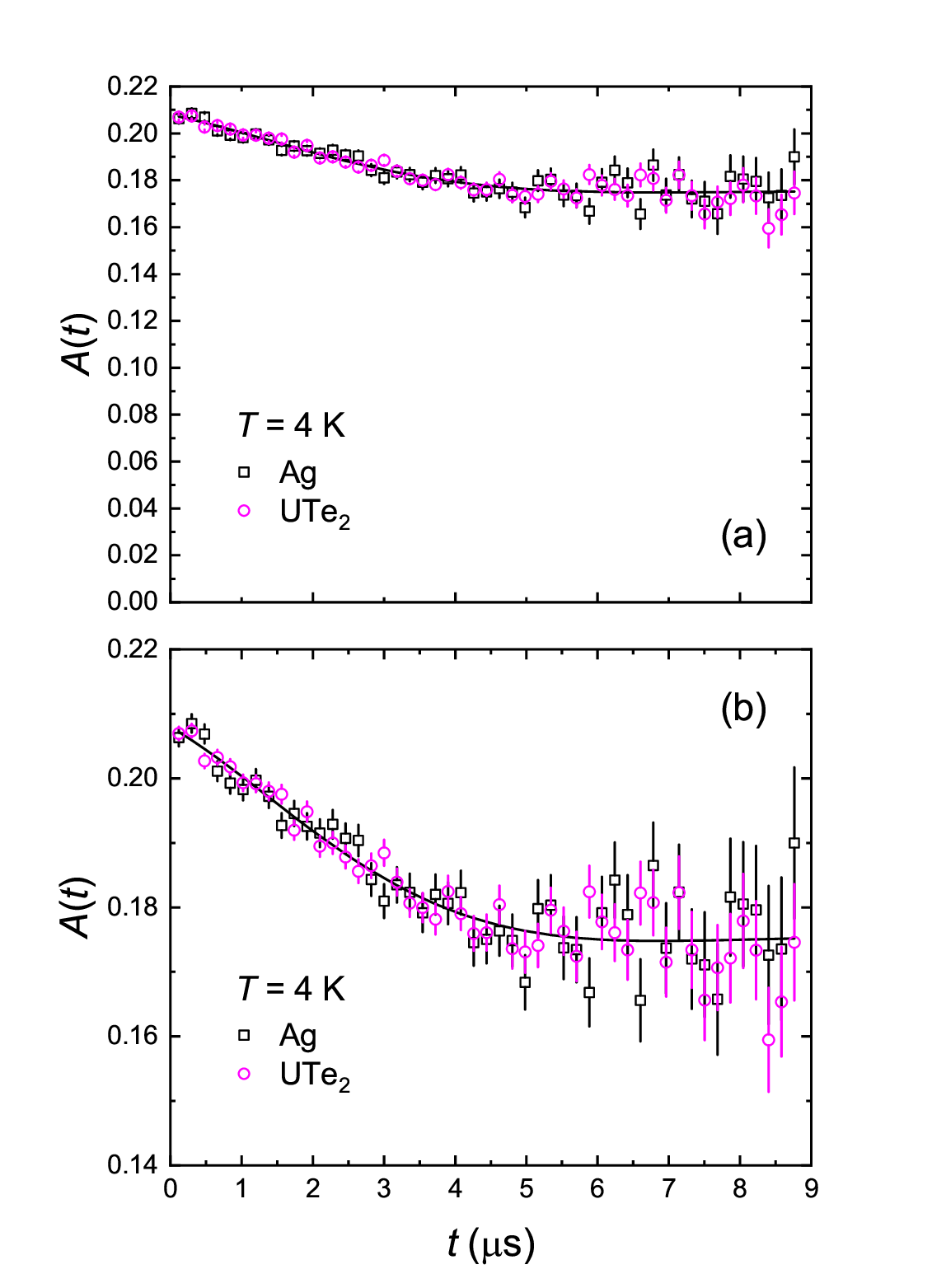}
\caption{(a) Comparison of the ZF-$\mu$SR asymmetry spectra for the Ag backing plate with (open circles) and without (open squares) the MSF-grown UTe$_2$ single crystals.
The solid curve is a fit of the ZF-$\mu$SR spectrum for the Ag plate to Eq.~(\ref{Eqn1}).
(b) Same as (a), but showing a zoom in on the vertical axis.}
\label{fig2}
\end{figure}  

Here we present results of a ZF-$\mu$SR study of MSF-grown UTe$_2$ single crystals. The single crystals all come from the same growth batch and
exhibit bulk superconductivity below $T_c \! = \! 2.01$~K, as determined from the midpoint of the specific heat jump shown in Fig.~\ref{fig1}(a). 
The crystals have an RRR value of 200 and the coefficient of the residual $T$-linear term in the specific heat below $T_c$ is approximately 9~mJ/mol$\cdot$K$^2$.
Figure~\ref{fig1}(b) shows a comparison of the temperature dependence of the bulk magnetic susceptibility ($\chi$) for a magnetic field of 1~kOe applied along the
three principal crystallographic axes. In contrast to the low-field behavior of $\chi(T)$ along the $a$-axis in CVT-grown UTe$_2$ single crystals 
\cite{Ran:19,Rosa:22}, $\chi_a(T)$ does not exhibit an upturn below $T \! \sim \! 10$~K.  

For the ZF-$\mu$SR measurements a mosaic of 24 single crystals was mounted on a $5 \! \times \! 17$~mm pure silver (Ag) backing plate 
thermally anchored to the Ag sample holder 
of an Oxford Instruments top-loading dilution refrigerator at the end of the M15 surface muon beamline at TRIUMF, Vancouver, Canada.
The MSF-grown UTe$_2$ single crystals covered 83~\% of the Ag backing plate.
The $c$ axis of each single crystal was aligned within $2^\circ$ of the normal of the Ag backing plate.
For the zero-field measurements, stray external magnetic fields at the sample position were reduced to $\lesssim \! 41$~mG using field compensation coils 
and the precession signal of muonium (Mu $\equiv \! \mu^+$e$^-$) in intrinsic Si as a sensitive magnetometer \cite{Morris:03}.
The ZF-$\mu$SR measurements were performed by implanting nearly 100~\% spin-polarized positive muons ($\mu^+$) into the sample 
with the initial muon spin polarization ${\bf P}(0)$ antiparallel to the muon beam direction (defined as the $z$-axis direction) and
parallel to the crystalline $c$-axis. 
The time evolution of the muon spin polarization $P_z(t)$ was determined
by detecting the muon decay positrons in a pair of opposing detectors positioned outside of the dilution refrigerator in front and behind the sample. 

\begin{figure}
\centering
\includegraphics[width=\columnwidth]{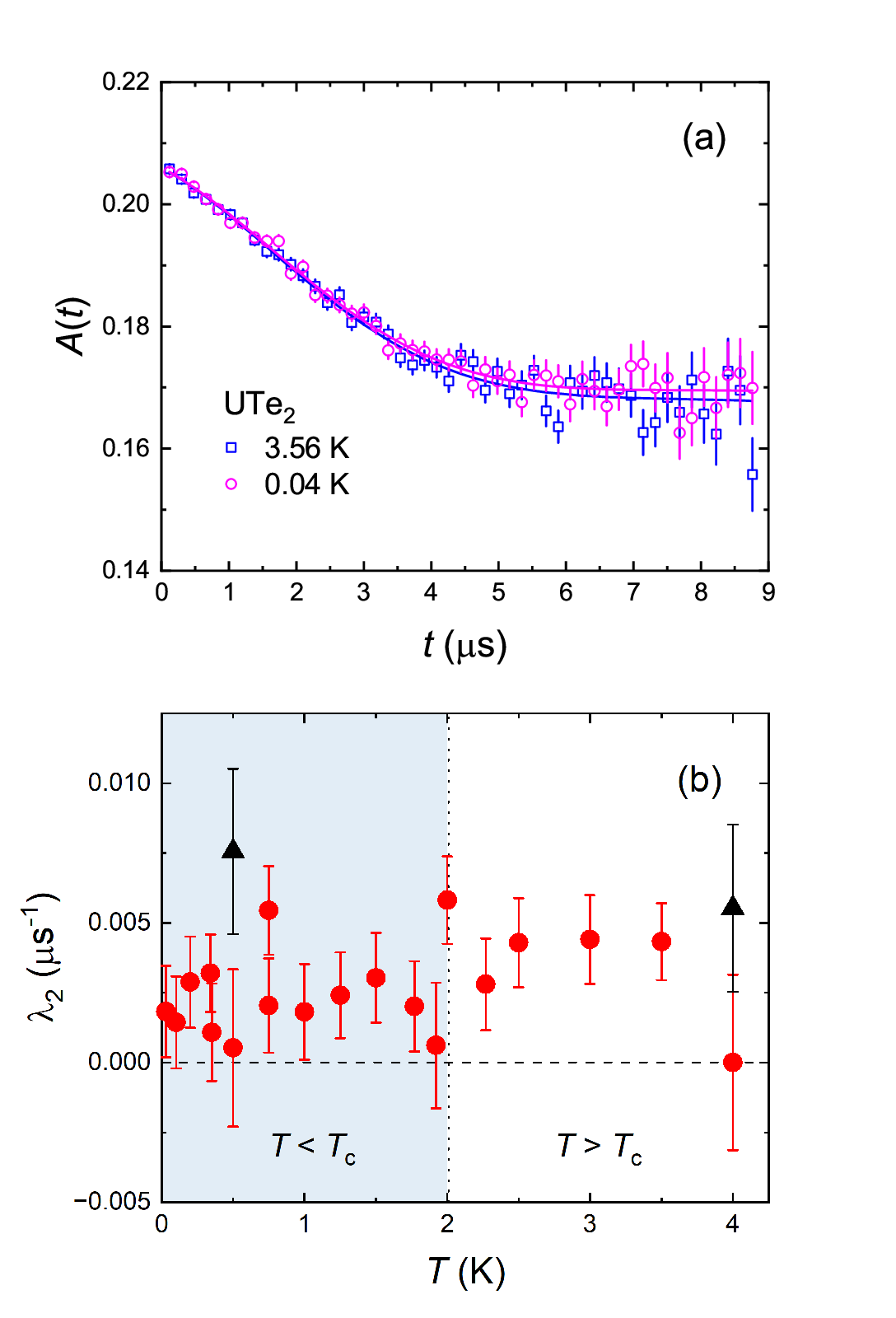}
\caption{(a) Comparison of the ZF-$\mu$SR asymmetry spectra recorded for the MSF-grown UTe$_2$ single crystals at $T \! = \! 3.56$~K and $T \! = \! 0.04$~K.
The solid curves are fits of the ZF-$\mu$SR spectra to Eq.~(\ref{Eqn1}) obtained from a global fit of the ZF-$\mu$SR spectra at all temperatures with $a_1$, $a_2$, $\lambda_1$ and
$\Delta$ as shared fitting parameters.
(b) Temperature dependence of $\lambda_2$ obtained from the global fit described in (a). The solid black triangles are values of $\lambda_2$ for the Ag backing plate without the sample
from fits assuming the same values of $\lambda_1$ and $\Delta$ generated by the global fit.}
\label{fig3}
\end{figure}
The ZF-$\mu$SR asymmetry spectra are well described by the following equation
\begin{eqnarray}
A(t) & = & a_0 P_z(t) \nonumber \\
& = & a_1 G_{z}^{\rm GKT}(\Delta, t) e^{-\lambda_1 t} + a_2 e^{-\lambda_2 t}   \, ,
\label{Eqn1}
\end{eqnarray}    
where $G_{z}^{\rm GKT}(\Delta, t)$ is a static Gaussian Kubo-Toyabe (GKT) function, characterized by the linewidth $\Delta/\gamma_{\mu}$ 
($\gamma_\mu/2 \pi$ is the muon gyromagnetic ratio) of a Gaussian distribution of local magnetic fields \cite{Hayano:79}.
Figure~\ref{fig2} shows a comparison of ZF-$\mu$SR asymmetry spectra for $T \! = \! 4$~K recorded for measurements of the Ag backing plate with and without the MSF-grown UTe$_2$ single crystals.
The observed relaxation of the ZF-$\mu$SR signal with time for the Ag plate without the sample is due to muons landing in material elsewhere within the dilution refrigerator.
This contribution is described by the first term in Eq.~({\ref{Eqn1}), which makes up $\sim \! 16$~\% of the total ZF-$\mu$SR signal ($a_1/a_0 \! \sim \! 16$~\%).
The remaining 84~\% of the ZF-$\mu$SR asymmetry spectrum is due to muons stopping in the Ag plate ($\sim \! 64$~\%) and a portion of the Ag sample holder ($\sim \! 20$~\%). 
This component of the ZF-$\mu$SR signal
is essentially non-relaxing ($\lambda_2 \! \sim \! 0$), as Ag does not possess electronic moments and has only very small nuclear moments
that do not cause an appreciable muon spin depolarization in the data time window. The relaxation of the ZF-$\mu$SR signal by randomly
oriented nuclear moments is also negligible in UTe$_2$, because the only stable uranium isotope with
non-zero nuclear spin is depleted $^{235}$U, which has a natural abundance of 0.20~\%, and the natural abundance of the tellurium isotopes with 
nuclear spin, $^{123}$Te and $^{125}$Te, are only 0.89~\% and 7~\%, respectively.
No discernible difference is observed between the ZF-$\mu$SR spectra for the Ag plate and the UTe$_2$ single crystals mounted on the Ag plate,
indicating there are no electronic moments fluctuating in the UTe$_2$ sample at this temperature that are slow enough to cause additional relaxation of the ZF-$\mu$SR signal.
With the sample in place, the non-relaxing component is due to muons stopping in the UTe$_2$ single crystals, in the Ag backing plate, and in a portion of the
Ag sample holder. These contributions are
separable in a transverse-field (TF) $\mu$SR measurement, due to a sizeable muon Knight shift in UTe$_2$ \cite{Azari:23}. 
The contribution of the UTe$_2$ sample to the non-relaxing part of the asymmetry spectrum is estimated to be at least 63~\% (or 53~\% of the total ZF-$\mu$SR signal)
from TF-$\mu$SR measurements for an applied magnetic field of 20~kOe.

The ZF-$\mu$SR asymmetry spectrum for the UTe$_2$ sample was recorded for 19 different temperatures between $T \! = \! 0.03$~K and $T \! = \! 4.0$~K. 
A global fit of the corresponding 19 different ZF-$\mu$SR spectra to Eq.~(\ref{Eqn1}) was carried out with the fit parameters $a_1$, $a_2$, $\lambda_1$ and $\Delta$ 
being shared parameters for all temperatures.
The global fit yielded the values $a_1/a_0 \! = \! 16.1(7)$~\%, $a_2/a_0 \! = \! 83.9(7)$~\%, 
$\lambda_1 \! = \!  0.158(7)$~$\mu$s$^{-1}$, and $\Delta \! = \!  0.280(7)$~$\mu$s$^{-1}$.
Representative fits to ZF-$\mu$SR asymmetry spectra included in the global fit are shown in Fig.~\ref{fig3}(a) and the temperature dependence of the exponential relaxation rate
$\lambda_2$ generated from the global fit is shown in Fig.~\ref{fig3}(b). There is no systematic increase in $\lambda_2$ with decreasing temperature, as previously
observed in CVT-grown single crystals \cite{Sundar:19,Sundar:23}. Hence, there are no electronic moments fluctuating slow enough to cause a detectable relaxation of the
ZF-$\mu$SR spectrum.
More importantly, there is no increase in $\lambda_2$ near $T_c$, and hence no evidence of spontaneous magnetic fields in the bulk associated
with a TRS broken superconducting state.        
 
Extremely slow intra U-ladder FM fluctuations along the $a$ axis have been inferred from NMR experiments on CVT-grown 
single crystals \cite{Tokunaga:22,Ambika:22,Fujibayashi:23}
that are far below the rate of spin fluctuations probed in inelastic neutron scattering studies \cite{Duan:20,Knafo:21}.
The low-energy FM spin fluctuations may be a consequence of vacancies in the U-ladder structure 
associated with a small U deficiency in some CVT-grown UTe$_2$ samples \cite{Sakai:22}. 
In particular, spins next to U vacancies may couple more strongly to the remaining neighboring spins \cite{Laukamp:98}, creating slowly fluctuating magnetic clusters 
that are detectable on the time scales of the NMR and $\mu$SR measurements.
However, it is clear from the current findings that the FM-like fluctuations observed in our initial ZF-$\mu$SR study of CVT-grown single crystals \cite{Sundar:19}
are not detectable in the absence of significant disorder. Consequently, it is an open question as to whether FM fluctuations are an intrinsic property of UTe$_2$.
 
To summarize, the main result of our ZF-$\mu$SR study of MSF-grown single crystals is the absence of spontaneous local magnetic fields in the bulk,
which are expected for a superconducting state that breaks TRS. This is in agreement with recent polar Kerr \cite{Ajeesh:23}, ultrasound \cite{Theuss:23} 
and NMR \cite{Matsumura:23} measurements that do not support a two-component superconducting order parameter.
 
\begin{acknowledgments}
We thank the personnel of the Centre for Molecular and Materials Science at TRIUMF for technical assistance with our $\mu$SR measurements.
J.E.S. and S.R.D. acknowledge support from the Natural Sciences and Engineering Research Council (NSERC) of 
Canada (PIN: 146772). S.S. acknowledges support from the Engineering and Physical Sciences Research Council (EPSRC) through grant EP/P024564/1.
Work at Los Alamos National Laboratory by S.M.T., J.D.T. and P.F.S.R. was supported by the U.S. Department of
Energy, and the Office of Basic Energy Sciences, Division of Materials Science. M.M.B. acknowledges support from the Laboratory Directed Research and Development program.
\end{acknowledgments}

\end{document}